# $Sn_{1-x}Bi_xO_2$ and $Sn_{1-x}Ta_xO_2$ (0 ≤ x ≤ 0.75): A first-principles study

## M.A. Ali, A.K.M.A. Islam[*]

*Department of Physics, University of Rajshahi, Rajshahi-6205, Bangladesh*

ABSTRACT

The structural, elastic, electronic and optical ($x = 0$) properties of doped $Sn_{1-x}Bi_xO_2$ and $Sn_{1-x}Ta_xO_2$ (0 ≤ $x$ ≤ 0.75) are studied by using the first-principles pseudopotential plane-wave method within the local density approximation. The independent elastic constants $C_{ij}$ and other elastic parameters of these compounds have been calculated for the first time. The mechanical stability of the compounds with different doping concentrations has also been studied. The electronic band structure and density of states are calculated and the effect of doping on these properties is also analyzed. It is seen that the band gap of the undoped compound narrowed with dopant concentration which disappeared for $x = 0.26$ for Bi doping and 0.36 for Ta doping. The materials thus become conductive oxides through the change in the electronic properties of the compound for $x$ ≤ 0.75 which may be useful for potential application. The calculated optical properties, e.g. dielectric function, refractive index, absorption spectrum, loss-function, reflectivity and conductivity of the undoped $SnO_2$ in two polarization directions are compared with both previous calculations and measurements.

## 1. Introduction

Transparent conducting oxide thin films are suitable for conducting solar window materials in thin film solar cells [1], heat reflectors for advanced glazing in solar applications [2,3] and as various gas sensors [4–7]. Tin oxide is the first transparent conductor to have received significant commercialization [8]. Among the different transparent conductive oxides, doped $SnO_2$ film seems to be the most appropriate for use in solar cells, owing to their low electrical resistivity and high optical transmittance. The high optical transmission, electrical conduction, and infrared reflection acquired by the films through doping have great potential for applications in optoelectronic devices, hybrid microelectronics, and photothermal conversion of solar energy [9]. Another interest is related to the fact that in most cases, impurities are included in order to improve the response of the material. In this sense, as an example, the interest in $SnO_2$ was recently renewed due to the discovery of high-temperature ferromagnetism in $Sn_{1-x}Co_xO_{2-\delta}$ films [10] with potential applications in spintronics. The impurities can introduce impurity levels in the gap of the semiconductor, modifying the electronic structure of the system. The description of these impurity levels (and their charge states) has attracted much interest. In general tin oxide $SnO_2$ is an n-type semiconductor with an energy gap of about ~3.6 eV [11]. The electronic properties of $SnO_2$ essentially change due to the introduction of 5A, B-groups elements. For example, $SnO_2$ films doped with Sb or Ta atoms (2–7 at %) exhibit a quasimetallic conductivity [12]. However a detailed understanding of the fundamental electronic structure and properties is required to obtain a high-quality material.

Investigations of undoped $SnO_2$, based on first-principles electronic structure calculations, have been previously reported in the literature [13-16]. Most of them dealt with the structural, phase transition, electronic properties, elastic behaviors, lattice dynamics properties, and phonon properties of undoped $SnO_2$. The electronic structure of doped $Sn_{1-x}M_xO_2$, (M=As, Sb, Bi, V, Nb, Ta) has been addressed by Zainullina [12] for $x = 0.0, 0.0625$ and 0.25 and electronic properties of only $Sn_{0.5}Sb_{0.5}O_2$ has been studied by Liang *et al.* [17]. But these studies seem to be inadequate in view of the fact that doping concentrations above 0.25 have not been considered. Extending the analysis to other doping concentrations may help in understanding the behavior of the materials for the doping ranges considered. Moreover the elastic properties of the doped materials are not available yet.

* Corresponding author: Tel: +88 0721 750980, Fax: +88 0721 750064
*Email address:* azi46@ru.ac.bd

Thus in this work, we will investigate the elastic and electronic properties of $Sn_{1-x}Bi_xO_2$ and $Sn_{1-x}Ta_xO_2$ for $0 \leq x \leq 0.75$. The optical properties of the undoped compound will also be included.

## 2. Method for calculation

The calculations are carried out utilizing the local density approximation (LDA) method with the Ceperley-Alder [18] form to describe the exchange and correlation potential based on the density functional theory (DFT) as implemented in the CASTEP code [19]. The stannum $5s^2$, $5p^2$ electrons and the oxygen $2s^2$, $2p^4$ electrons are treated as part of the valence states. The Monkhorst-Pack scheme *k*-point sampling has been used for integration over the first Brillouin zone [20]. The plane-wave cutoff energy is 1300 eV. The convergence criteria for structure optimization and energy calculation were set to ultrafine quality with the *k*-point mesh of $5 \times 5 \times 8$ for the crystal structure, which make the tolerance for self-consistent field, energy, maximum force, maximum displacement, and maximum stress to be $5.0 \times 10^{-7}$ eV/atom, $5.0 \times 10^{-6}$ eV/atom, 0.01 eV/Å, $5.0 \times 10^{-4}$ Å, and 0.02 GPa, respectively. The Broyden–Fletcher–Goldfarb-Shenno (BFGS) [21] minimization technique has been used for the structural parameters of tetragonal $SnO_2$, $Sn_{1-x}Bi_xO_2$ and $Sn_{1-x}Ta_xO_2$.

## 3. Results and discussion

### 3.1. Structural properties

The rutile-type $SnO_2$ with space group $P4_2/mnm$ and symmetry $D_{4h}^{14}$ belongs to the tetragonal system which is shown in Fig. 1. The unit cell contains two molecules with the tin atoms at the positions (0, 0, 0) and (0.5, 0.5, 0.5) and the oxygen atoms at $\pm$ ($u$, $u$, 0), ((1/2+$u$, 1/2–$u$, 1/2) (1/2–$u$, ½+$u$, 1/2) with $u$ = 0.306 [22]. Each Sn atom is in the central site of an octahedron which formed by four rectangular basal O atoms ($O_1$) and two vertex O atoms ($O_2$). The total energy is minimized by the geometry optimization and the optimized values of structural parameters of $SnO_2$, $Sn_{1-x}Bi_xO_2$ and $Sn_{1-x}Ta_xO_2$ are shown in Table 1 along with other available theoretical and experimental results.

**Fig. 1.** Unit cell of $SnO_2$.

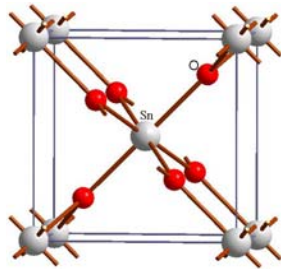

### 3.2. Elastic properties

The independent elastic constants of $Sn_{1-x}Bi_xO_2$ and $Sn_{1-x}Ta_xO_2$ are calculated under zero external pressure and these are shown in Table 2. Our six independent elastic constants for $SnO_2$ are compared with the other experimental calculation [24] in Table 2. To our knowledge, the elastic constants $Sn_{1-x}Bi_xO_2$ and $Sn_{1-x}Ta_xO_2$ are calculated for the first time in this study. The independent elastic constants are all positive and satisfy the well-known Born's criteria for stability of tetragonal crystals: $C_{11} > 0$, $C_{33} > 0$, $C_{44} > 0$, $C_{66} > 0$, $(C_{11} - C_{12}) > 0$, $(C_{11} + C_{33} - 2C_{13}) > 0$ and $[2(C_{11} + C_{12}) + C_{33} + 4C_{13}] > 0$.

Table 1   The optimized structural parameters of $Sn_{1-x}Bi_xO_2$ and $Sn_{1-x}Ta_xO_2$.

| Compound | $a = b$ (Å) | $c$ (Å) | $c/a$ | $V$ (Å³) |
|---|---|---|---|---|
| $SnO_2$ | 4.673 | 3.148 | 0.672 | 68.76 |
| | 4.737[†] | 3.186[†] | 0.672 | 71.49 |
| | 4.888[δ] | 3.284[δ] | 0.672 | 78.44 |
| | 4.695[‡] | 3.160[‡] | 0.672 | 68.76 |
| $Sn_{0.75}Bi_{0.25}O_2$ | 4.753 | 3.213 | 0.675 | 72.58 |
| $Sn_{0.5}Bi_{0.5}O_2$ | 4.815 | 3.287 | 0.681 | 76.21 |
| $Sn_{0.25}Bi_{0.75}O_2$ | 4.869 | 3.347 | 0.687 | 79.33 |
| $Sn_{0.75}Ta_{0.25}O_2$ | 4.836 | 3.287 | 0.679 | 76.86 |
| $Sn_{0.5}Ta_{0.5}O_2$ | 4.997 | 3.478 | 0.695 | 86.86 |
| $Sn_{0.25}Ta_{0.75}O_2$ | 5.039 | 3.568 | 0.707 | 90.60 |

[†]Expt. [23], [δ]Ref. [13], [‡]Ref. [15].

Table 2   Calculated independent elastic constants in GPa of $Sn_{1-x}Bi_xO_2$ and $Sn_{1-x}Ta_xO_2$.

| Compound | $C_{11}$ | $C_{12}$ | $C_{33}$ | $C_{44}$ | $C_{66}$ | $C_{13}$ |
|---|---|---|---|---|---|---|
| $SnO_2$ | 306.3 | 212.6 | 516.0 | 114.7 | 245.2 | 179.2 |
| | 261.7[a] | 177.2[a] | 449.6[a] | 103.7[a] | 207.4[a] | 155.5[a] |
| $Sn_{0.75}Bi_{0.25}O_2$ | 266.9 | 189.8 | 473.7 | 126.3 | 204.1 | 163.8 |
| $Sn_{0.5}Bi_{0.5}O_2$ | 233.1 | 170.5 | 399.5 | 113.2 | 151.7 | 153.6 |
| $Sn_{0.25}Bi_{0.75}O_2$ | 212.4 | 160.2 | 301.0 | 96.0 | 109.8 | 158.4 |
| $Sn_{0.75}Ta_{0.25}O_2$ | 256.7 | 192.1 | 437.7 | 98.8 | 200.1 | 164.1 |
| $Sn_{0.5}Ta_{0.5}O_2$ | 169.8 | 135.8 | 312.5 | 90.5 | 146.5 | 112.5 |
| $Sn_{0.25}Ta_{0.75}O_2$ | 182.1 | 141.4 | 406.2 | 80.9 | 152.8 | 133.9 |

[a]Expt. [24].

The calculated elastic parameters (bulk moduli $B$, compressibility $K$, shear moduli $G$, Young's moduli $Y$ and the Poisson ratio $v$) are given in Table 3. $Y$ and $v$ are computed using the relationships: $Y = 9BG/(3B + G)$, $v = (3B - Y)/6B$ [25]. Here Hill approximation is used for all calculations. Hill [26] proved that the Voigt and Reuss equations represent upper and lower limits of the true polycrystalline constants. He showed that the polycrystalline moduli are the arithmetic mean values of the moduli in the Voigt ($B_V$, $G_V$) and Reuss ($B_R$, $G_R$) approximation, and are thus given by $B_H \equiv B = \frac{1}{2}(B_R + B_V)$ and $G_H \equiv G = \frac{1}{2}(G_R + G_V)$, where $B$ and $G$ represent the bulk modulus and shear modulus respectively. The expression for Reuss and Voigt moduli can be found elsewhere [27]. The tetragonal shear moduli, $G' = (C_{11} - C_{12})/2$ and Zenger's anisotropy index, $A = 2C_{44}/(C_{11} - C_{12})$ [28] are calculated.

For each of the cases, we see that $B > G > G'$ which implies that the limiting parameter for mechanical stability of these materials is the tetragonal shear modulus $G'$. The Young's modulus $Y$ measures the response to an uniaxial stress averaged over all directions and is used often to denote a measure of stiffness. From our calculations the undoped $SnO_2$ is stiffest than the doped $Sn_{1-x}Bi_xO_2$ and $Sn_{1-x}Ta_xO_2$. The deviations of Zenger's anisotropy from unity measure the degree of elastic anisotropy. Our results show that all the compounds under consideration are highly anisotropic. As the Poisson's ratio for brittle materials is small, whereas for ductile metallic materials $v$ is typically 0.33 [29], we can see that the examined doped compounds with $x > 0.25$ show ductile metallic behavior. Further we note that one of the most widely used malleability indicator of materials is Pugh ductility index (G/B) [30]. If $G/B$<0.5, the material will behave as ductile material. On the other hand, if $G/B$> 0.5, the material will show brittleness. According to this criterion also all the compounds possess ductile behavior.

Table 3  Calculated elastic parameters of polycrystalline $SnO_2$, $Sn_{1-x}Bi_xO_2$ and $Sn_{1-x}Ta_xO_2$.

| Compound | $B$ (GPa) | $K$ (1/GPa) | $G$ (GPa) | $G'$ (GPa) | $Y$ (GPa) | $v$ | $A$ |
|---|---|---|---|---|---|---|---|
| $SnO_2$ | 248.2 | 0.0041 | 115.7 | 46.8 | 300.4 | 0.20 | 2.42 |
|  | 212.3[a] | 0.0047* | 101.8[a] | 42.2* | 263.3* | 0.29* | 2.45* |
| $Sn_{0.75}Bi_{0.25}O_2$ | 217.3 | 0.0046 | 90.8 | 38.6 | 239.2 | 0.31 | 3.27 |
| $Sn_{0.5}Bi_{0.5}O_2$ | 194.0 | 0.0051 | 74.8 | 31.3 | 199.0 | 0.32 | 3.61 |
| $Sn_{0.25}Bi_{0.75}O_2$ | 181.7 | 0.0055 | 57.00 | 26.1 | 154.8 | 0.35 | 3.67 |
| $Sn_{0.75}Ta_{0.25}O_2$ | 213.5 | 0.0046 | 76.32 | 32.3 | 204.6 | 0.34 | 3.13 |
| $Sn_{0.5}Ta_{0.5}O_2$ | 146.0 | 0.0068 | 49.5 | 34.0 | 133.5 | 0.34 | 5.33 |
| $Sn_{0.25}Ta_{0.75}O_2$ | 159.2 | 0.0062 | 54.5 | 20.3 | 147.1 | 0.34 | 3.98 |

[a]Expt. [24]. (*) marked values are calculated by us using experimental elastic constants.

### 3.3. Electronic properties

The band structures of $SnO_2$, $Sn_{1-x}Bi_xO_2$ and $Sn_{1-x}Ta_xO_2$ along high-symmetry directions of the crystal Brillouin zone (BZ) are displayed in Figs. 2 and 3. The Fermi level is chosen to be zero of the energy scale. In the case of $SnO_2$, a direct band gap of 1.38 eV at $\Gamma$ is found [Fig. 2 (a)]. This may be compared with the values of 1.38 eV and 1.7 eV found by Liu *et al.* [13] using LDA as implemented in CASTEP and by Errico [14] using the FP-LAPW method with relativistic corrections, respectively. These values are much smaller than the experimental value [11]. The underestimation of the band gap is a well-known effect of the DFT calculation and can be corrected by several approaches. One way is to identify the Kohn-Sham eigenvalues with quasi-particle energies. This can be accounted for by a rigid shift of the conduction band upward with respect to the valence band [31, 32]. Moreover the phonons and their optical effects have also not been taken into account. Despite this limitation, DFT is able to reproduce trends, such as a variation in the band gap due to changes for different doping concentration.

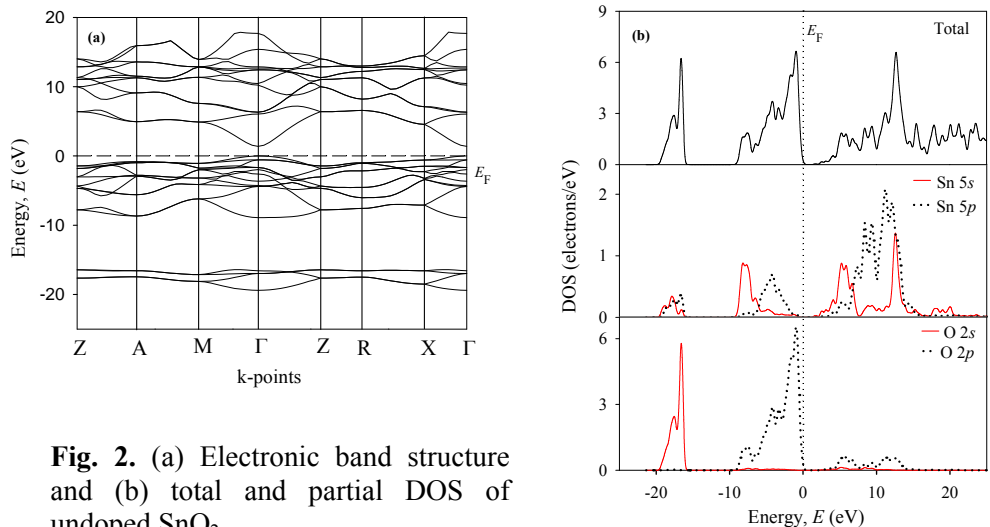

**Fig. 2.** (a) Electronic band structure and (b) total and partial DOS of undoped $SnO_2$.

The corresponding total and partial density of states are shown in Fig. 2 (b). The band gap reduces to zero at $x = 0.26$ for Bi substitution and 0.36 for Ta substitution (figures not shown). In the case of $SnO_2$, the band structures around Fermi level, $E_F$ are derived mainly from O-$2p$ and Sn-$5p$ orbitals. The contribution from Sn-$5p$ orbitals is noticeable but an order of magnitude smaller than the contribution from O-$2p$ orbitals. In the valence bands (VBs) of the compounds, O-$2s$ states are dominant below -15 eV with a slight contribution of the Sn-$5s$ and Sn-$5p$ states. In the range between -10 and 0 eV, the O-$2p$ states are the most dominant, especially at the region closer to the Fermi level ($E_F$), while a hybridization between Sn-$5s$, Sn-$5p$ and O-$2p$ states could be seen in the range of -10 eV. The density of states of the conduction bands (CBs) are mainly derived from Sn-$5s$ and Sn-$5p$ orbitals while O-$2p$ states also have a little contribution. In the bottom of the CBs, the hybridization between Sn-$5s$ and O-$2p$ states also observed.

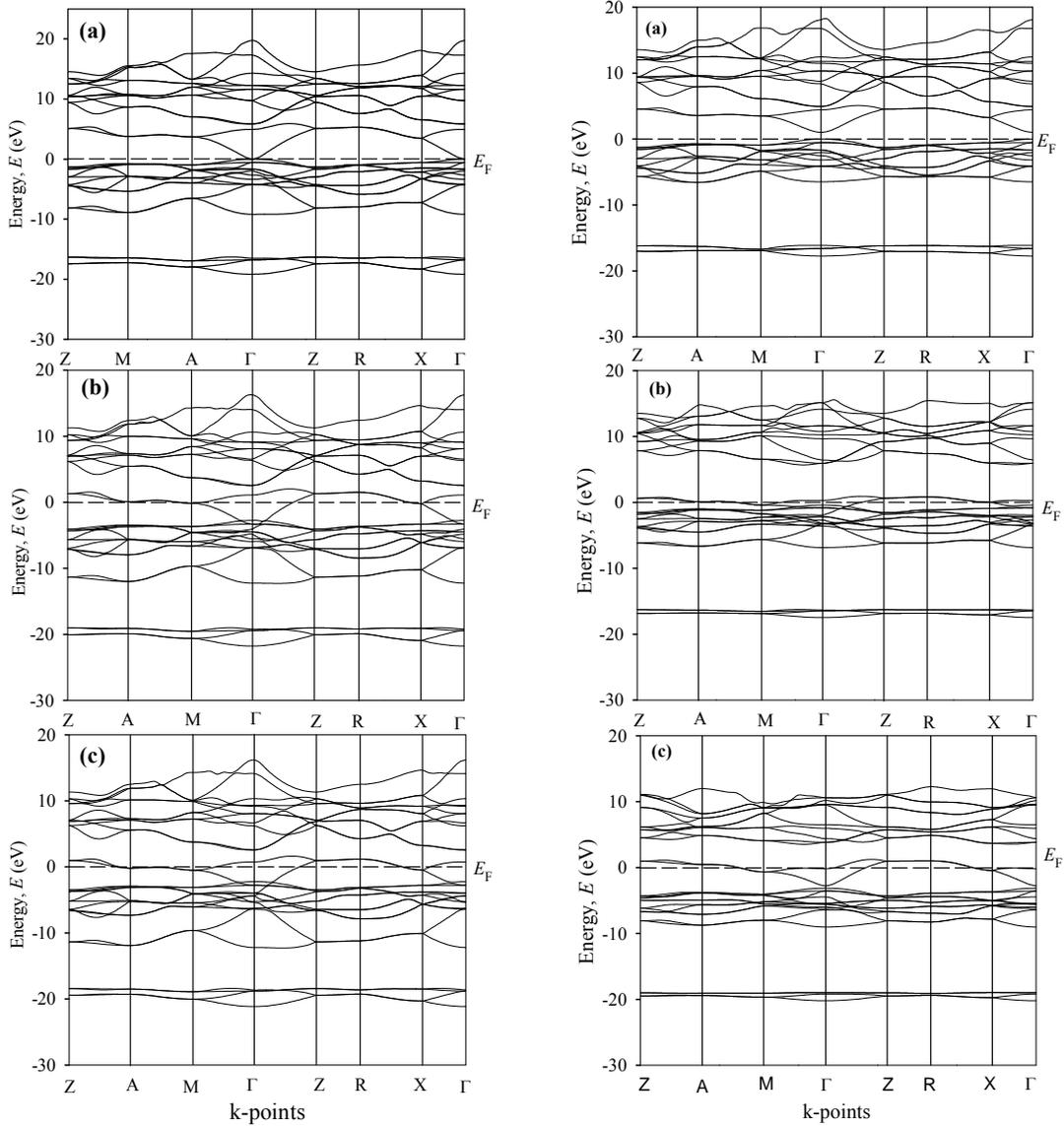

**Fig. 3.** Electronic band structure of $Sn_{1-x}Bi_xO_2$ (*left panel*) and $Sn_{1-x}Ta_xO_2$ (*right panel*) for (a) $x = 0.25$, (b) $x = 0.5$, and (c) $x = 0.75$, respectively.

In order to study the doping effects on electronic properties due to substitution of Bi/Ta on the Sn sub-lattice we now refer to the band structures and density of states of $Sn_{1-x}Bi_xO_2$ and $Sn_{1-x}Ta_xO_2$ for various values of doping levels as shown in Figs. 3 and 4. In comparison to band structure for $x = 0$ [Fig. 2 (a)], there are a number of additional features related to the effects due to the substitution of Bi/Ta. The main feature of the band spectra of the doped compounds (Figs. 3) is the appearance of additional bands in the gap region which can clearly be seen in these figures.

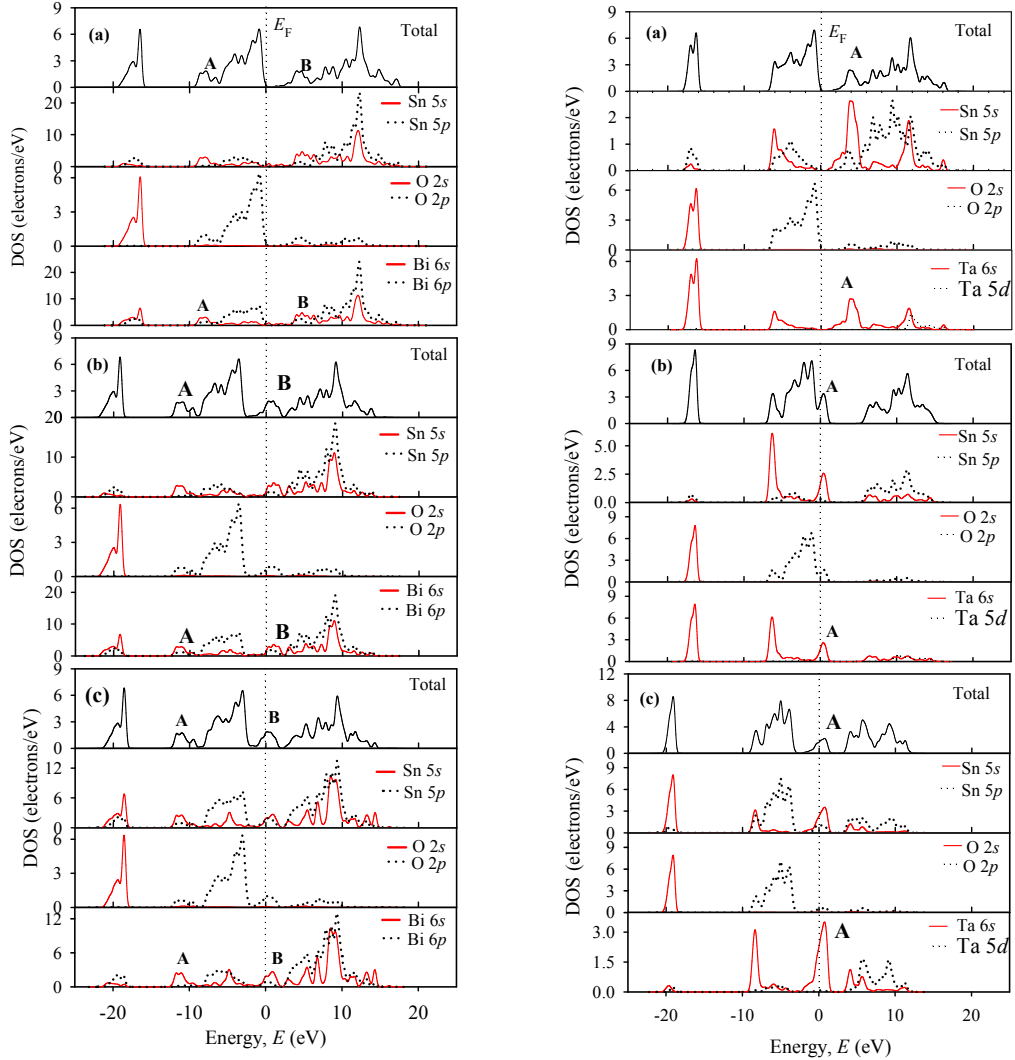

**Fig. 4.** Total and partial DOS of $Sn_{1-x}Bi_xO_2$ (*left panel*) and $Sn_{1-x}Ta_xO_2$ (*right panel*) for (a) $x = 0.25$, *(*b) $x = 0.5$, and (c) $x = 0.75$, respectively.

Now let us have a closer look on the effect of doping for $x \geq 0.25$. In Fig. 4 for $Sn_{1-x}Bi_xO_2$ there are two additional **A** and **B** bands consisting of the valence $ns$-states of Bi atoms with the addition of the O-$2p$ and Sn-$5s$ states. The **A** band is at the bottom of the valence band and the **B** band is at the bottom of conduction band for $Sn_{0.75}Bi_{0.25}O_2$, but **B** band lies in the energy gap for $Sn_{0.5}Bi_{0.5}O_2$ and $Sn_{0.25}Bi_{0.75}O_2$. Moreover, there is difference in the positions of **A** and **B** bands in the band structure of $Sn_{0.75}Bi_{0.25}O_2$, $Sn_{0.5}Bi_{0.5}O_2$ and $Sn_{0.25}Bi_{0.75}O_2$. The positions of **A** band are at -0.785, -11.237 and -10.979 eV and **B** band are at 3.208, 0.668 and 0.187 eV for $Sn_{0.75}Bi_{0.25}O_2$, $Sn0.5Bi_{0.5}O_2$ and $Sn_{0.25}Bi_{0.75}O_2$, respectively. The position of **B** band is shifted towards the Fermi level with the increase of doping level with subsequent increase in conductivity. In the case of $Sn_{1-x}Ta_xO_2$ for all the doping levels considered there is only one additional band **A** in the bottom of the conduction band. The positions of the band **A** are located at 3.7, 0.462 and 0.781eV for $x$ = 0.25, 0.5 and 0.75, respectively. We see that the band gap narrowed and reduces to zero as doping increases. In case of $Sn_{1-x}Bi_xO_2$ and $Sn_{1-x}Ta_xO_2$, metallic conductivity starts at 26% for Bi doping and at 36 % for Ta doping.

As can be seen in Figs. 3 and 4 the band structure and DOS of $Sn_{0.5}Bi_{0.5}O_2$, $Sn_{0.25}Bi_{0.75}O_2$, $Sn_{0.5}Ta_{0.5}O_2$, and $Sn_{0.25}Ta_{0.75}O_2$ show increased metallicity for these compounds. The Fermi level crosses dispersion curve of the Bi/Ta 6s state which overlaps with the conduction band. The introduction of the Bi or Ta impurities caused Bi or Ta 6s distribution of electronic states from Fermi level to the bottom of the conduction band. The activated conductivity of these compounds is determined by the transition of the electrons from the donor band to the conduction band. The value of the DOS for x= 0.25 [Fig. 4 (a,b,c)] also increased progressively as doping level of Bi/Ta atom is increased. It is known that the metallic conductivity is proportional to the density of states at Fermi level. Therefore the increase of DOS will bring the growth of conductivity. The envelope of the DOS shifted towards the low energy direction by 2.5 eV, 1 eV and 1 eV for $Sn_{0.5}Bi_{0.5}O_2$, $Sn_{0.25}Bi_{0.75}O_2$ and $Sn_{0.25}Ta_{0.75}O_2$, respectively.

### 3.4. Optical properties

The optical properties of $SnO_2$ may be obtained from the complex dielectric function, $\varepsilon_1(\omega) = \varepsilon_1(\omega) + i\varepsilon_2(\omega)$. The imaginary part $\varepsilon_2(\omega)$ is obtained from the momentum matrix elements between the occupied and the unoccupied electronic states and calculated directly by CASTEP [31] using:

$$\varepsilon_2(\omega) = \frac{2e^2\pi}{\Omega\varepsilon_0} \sum_{k,v,c} \left| \psi_k^c \left| u.r \right| \psi_k^v \right|^2 \delta\left(E_k^c - E_k^v - E\right) \tag{1}$$

where $u$ is the vector defining the polarization of the incident electric field. $\omega$ is the light frequency, $e$ is the electronic charge and $\psi_k^c$ and $\psi_k^v$ are the conduction and valence band wave functions at $k$, respectively. The real part is derived from the imaginary part $\varepsilon_2(\omega)$ by the Kramers-Kronig transform. All other optical constants, such as refractive index, absorption spectrum, loss-function, reflectivity and conductivity (real part) are those given by Eqs. 49 to 54 in ref. [31].

The optical properties (eight constants) of the undoped $SnO_2$ from the polarization vectors (1 0 0) and (0 0 1) are shown in Fig. 5 for energy range up to ~50 eV. Some recent calculations are available on some of these properties [13, 33, 34]. Results on $\varepsilon_1$, $\varepsilon_2$, $n$, $k$ of one such calculations due to Liu *et al*. [13] in (1 0 0) polarization direction are compared with our results in Fig. 5 (a-d). The curves seem to be similar but not identical. There are some differences in the positions of the peaks in the two calculations.

The probability of photon absorption is directly related to the imaginary part of the optical dielectric function $\varepsilon(\omega)$. In the case of $\varepsilon_2(\omega)$, we find that there are three main peaks at 7.6, 10 and 14.7 eV. The three peaks due to Liu *et al*. [13] are slightly shifted towards higher photon energy. The experimental data [35] show the first two peaks, but the third one at ~ 15 eV is hardly visible. The peak at 7.6 eV mainly arises from the electron transition from the O-2$p$ orbitals to Sn-5$p$ orbitals. The electron transition from the hybridization orbitals of O-2$p$ and Sn-5$p$ to the hybridization orbitals of Sn-5$s$ and Sn-5$p$ may lead the peaks at 10 and 14.7 eV. For the real part $\varepsilon_1(\omega)$ of $SnO_2$, [Fig. 5 (a), left panel] the first main peak is at 1.6 eV. The negative value of $\varepsilon_1$ between 14.7 to 23 eV indicates that the material shows metallic properties in this range of photon energy.

The refractive index and the extinction coefficient are displayed in Fig. 5 (c, d) (left panel). The static refractive index of $SnO_2$ is found to have the values 2.15 and 2.1 for the two polarization vectors (0 0 1) and (1 0 0), respectively. These values compare well with 1.997 and 2, for the bulk and the values 1.8 – 1.9 for MLD films [36]. This shows that the vibrational contribution to $\varepsilon_1$ is the origin of high-permittivity feature of SnO$_2$. Further following Borges *et al.* [34] and their evaluation of the TO frequencies, it may be remarked here that low frequency optical phonons are the primary contributors to the dielectric tensor.

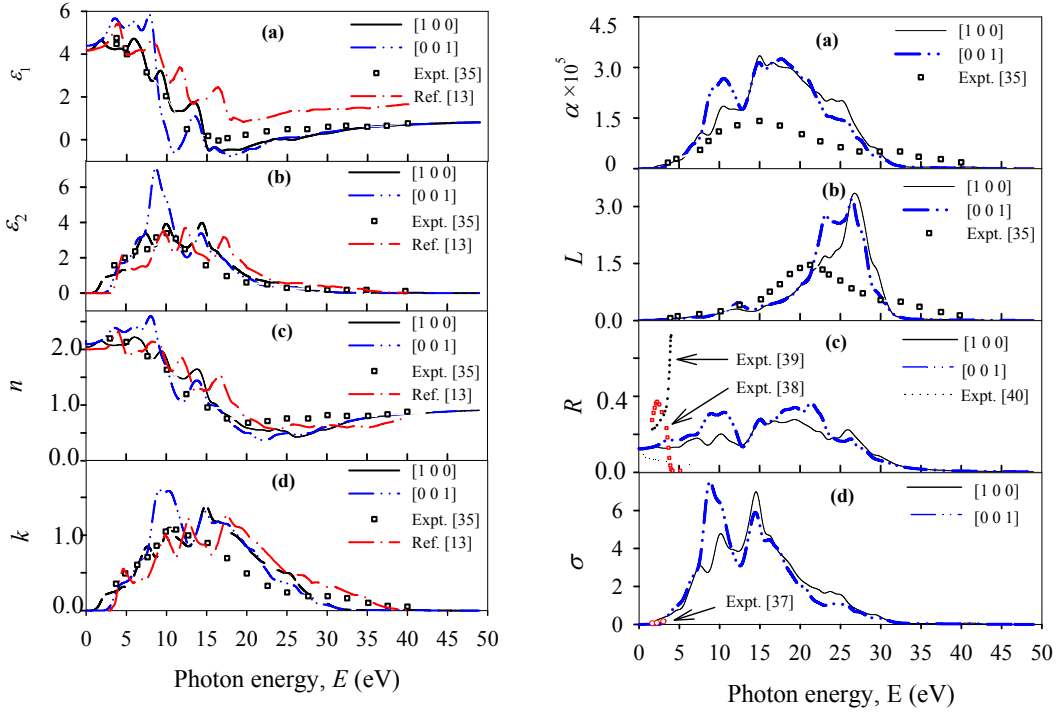

**Fig. 5.** *Left panels*: Energy dependent (a) real part of dielectric function, (b) imaginary part of dielectric function, (c) refractive index, and (d) extinction coefficient of undoped SnO$_2$ for two polarization directions. *Right panels*: Energy dependent (a) absorption, (b) loss function, (c) reflectivity, and (d) real part of conductivity of undoped SnO$_2$ for two polarization directions. Results of other calculation [13] and measurements [35, 38-40] are also shown.

The absorption coefficient displayed in Fig. 5 (a) (right panel) shows that the structure has no absorption band at low energy range due to its semiconducting nature. We see that the absorption spectrum starts at ~1.4 eV. Figs. 5 (b) (right panel) shows the loss function $L(\omega)$ which is an important parameter describing the energy loss of a fast electron traversing in the material. The peaks represent the characteristics associated with the plasma resonance and the corresponding frequency is the so-called plasma frequency (14.7 eV). The material exhibits dielectric [$\varepsilon_1(\omega) > 0$] and metallic [$\varepsilon_1(\omega) < 0$] behaviors above and below plasma frequency, respectively. The reflectivity is 0.12 to 0.18 for photon energy of 0 to 10 eV [Fig. 5(c): right panel] which indicates that SnO$_2$ material is mostly transmitting in this region. A photon energy of ~ 1.4 eV initiates photoconductivity as can be seen in Fig. 5 (d) (right panel).

Experimental results on six constants $\varepsilon_1$, $\varepsilon_2$, $n$, $k$, $\alpha$ and $L$ have been obtained using reflection electron energy loss spectroscopy (REELS) measurements between 4 and 40 eV of SnO$_2$ thin films [35]. The present theoretical curves of $\varepsilon_1$, $\varepsilon_2$, $n$, and $k$ are in overall agreement with the experimental data as shown in the left panels of Fig. 5 (a-d). The calculated absorption and loss function shown on the right panels of Fig. 5 follow the experimental values reasonably up to ~15 eV. Based on spectroscopic ellipsometry (SE) measurements optical and structural properties of the $SnO_2$ films have been obtained by Anastasescu *et al.* [37] upto ~ 4 eV. The calculated real part of conductivity (imaginary part, not shown, is either zero or small for energy < 2 eV) data fit well with this measurement. To our knowledge, we did not find suitable experimental data of reflectivity for proper comparison because the available data are for very low energy (< 3 eV) [38-40], and moreover all these data originate from samples as thin and thick films with different substrates, films with differences in methods of preparation (resulting in porosity and defect, to some extent) and measurement techniques and thus seen to vary widely from each other. It is also necessary to take account for the dependence of reflectivity on film thickness. So a meaningful comparison is hardly possible at this time.

## 4. Conclusions

The structural, elastic and electronic properties of $Sn_{1-x}Bi_xO_2$ and $Sn_{1-x}Ta_xO_2$ are calculated using first-principles method. The analysis of elastic constants shows that $Sn_{1-x}Bi_xO_2$ and $Sn_{1-x}Ta_xO_2$ are mechanically stable and anisotropic for $0 \leq x \leq 0.75$. The effect of Bi and Ta doping on the electronic properties are also analyzed. The main feature of the band spectra of the doped compounds is the creation of additional bands in the gap region. We see that the band gap value of the undoped compound narrowed due to doping. The materials $Sn_{1-x}Bi_xO_2$ and $Sn_{1-x}Ta_xO_2$ become conductive oxides for 26% Bi substitution and 36 % Ta substitution, respectively. Further, the optical properties, e.g. dielectric function, refractive index, absorption spectrum, loss-function, reflectivity and conductivity of undoped $SnO_2$ analyzed and compared with available theoretical and experimental results.